\documentclass[12pt]{article}
\usepackage{graphicx}
\usepackage{psfrag}
\usepackage[round]{natbib}
\usepackage{here}
\usepackage{amsmath}
\usepackage{caption} 
\usepackage{subcaption} 
\usepackage{microtype} 
\usepackage{hyperref}

\def \IR{\hbox{{\rm I}\kern-.2em\hbox{{\rm R}}}}
\title{Adaptive Geostatistical Design and Analysis for Sequential Prevalence Surveys}
\author{Michael G. Chipeta$^{a,b,c}$, Dianne J. Terlouw$^{a, c, d}$, Kamija Phiri$^a$ and Peter J. Diggle$^b$ \\ \\
$^a$College of Medicine, University of Malawi, Blantyre, Malawi\\
$^b$Lancaster Medical School, Lancaster University, Lancaster, UK\\
$^c$Malawi-Liverpool Wellcome Trust Clinical Research Programme, Blantyre, Malawi\\
$^d$Liverpool School of Tropical Medicine, Liverpool, UK}
\begin{document}

\maketitle \vspace{1cm}
{\bf Abstract}\\
Non-adaptive geostatistical designs (NAGD) offer standard ways of collecting and analysing geostatistical data in which sampling locations are fixed in advance of any data collection. In contrast, adaptive geostatistical designs (AGD) allow collection of exposure and outcome data over time to depend on information obtained from previous information to optimise data collection towards the analysis objective.AGDs are becoming more important in spatial mapping, particularly in poor resource settings where uniformly precise mapping may be unrealistically costly and priority is often to identify critical areas where interventions can have the most health impact. Two constructions are: \emph{singleton} and \emph{batch} adaptive sampling. In singleton sampling, locations $x_i$ are chosen sequentially and at each stage, $x_{k+1}$ depends on data obtained at locations $x_1,\ldots, x_k$. In batch sampling, locations are chosen in batches of size $b > 1$, allowing new batch, $\{x_{(k+1)},\ldots,x_{(k+b)}\}$, to depend on data obtained at locations $x_1,\ldots,x_{kb}$. In most settings, batch sampling is more realistic than singleton sampling. We propose specific batch AGDs and assess their efficiency relative to their singleton adaptive and non-adaptive counterparts by using simulations.We show how we apply these findings to inform an AGD of a rolling Malaria Indicator Survey, part of a large-scale,  five-year  malaria transmission reduction project in Malawi.

{\bf Keywords.} Adaptive sampling strategies, Spatial statistics, Geostatistics, Malaria, Prevalence mapping

\section{Introduction}\label{sec:introduction}

Geostatistics has its origins in the South African mining industry \citep{Krige1951}, and was subsequently developed by Georges Matheron and colleagues into a self-contained methodology for solving prediction problems arising principally in mineral exploration; \citet{Chiles2012} is a recent book-length account. Within the general statistics research community, the term geostatistics more generally refers to the branch of spatial statistics that is concerned with investigating an unobserved  spatial  phenomenon $S = \{S(x): x \in D \subset \IR^2\}$ , where $D$ is a geographical region of interest, using data  in the form of measurements $y_i$  at locations $x_i \in D$. Typically, each  $y_i$   can  be  regarded  as  a  noisy  version  of $S(x_i)$.  We  write  $X  = \{x_1 , \ldots , x_n\}$ and call $X$ the {\it sampling  design}.

Geostatistical analysis can address either or both of two broad objectives: {\it estimation} of the  parameters that define a stochastic model  for the unobserved process $S$  and  the  observed data  $\{(y_i , x_i ) : i = 1, ..., n\}$; {\it prediction}  of the  unobserved  realisation  of $S(x)$ throughout $D$, or particular
characteristics of this realisation, for example its average value.
	
A key consideration for geostatistical design is that sampling designs that are efficient for parameter estimation are generally inefficient for prediction, and vice versa. Since parameter values are always unknown in practice, design for prediction therefore involves a compromise.  Furthermore, the diversity of potential  predictive targets requires design strategies to be context-specific. Another important distinction is between {\it non-adaptive} sampling designs that must be completely specified prior to data-collection, and {\it adaptive} designs, for which data are collected over a period of time and later sampling locations can depend on data collected from earlier locations.

In this paper we formulate, and evaluate through simulation studies, a class of adaptive design strategies that address two compromises:  between efficient parameter estimation and efficient prediction; and between theoretical advantages and practical constraints. The motivation for our work is the mapping of malaria prevalence in rural communities through a series of ``rolling malaria indicator surveys," henceforth rMIS \citep*{Roca-Feltrer2012}. Malaria prevalence is highly heterogenous in time and space.  Adaptive design is especially relevant here because resource constraints make it difficult to achieve uniformly precise predictions throughout the region of interest. Hence, as data accrue over the study-region $D$ it becomes appropriate to focus progressively on sub-regions of $D$ where precise prediction is needed to inform public health action, for example to prioritise sub-regions for early intervention.

In Section \ref{sec:methods} we review the existing  literature on adaptive geostatistical design and set out the methodological framework  within  which we will specify and evaluate adaptive design strategies. Section \ref{sec:framework} describes our proposed class of adaptive designs for efficient prediction. Section  \ref{sec:simulation}  gives the results of a simulation  study in which we compare the predictive efficiency of our proposed design strategy with simpler, non-adaptive strategies. Section \ref{sec:application} is an application to the design of an ongoing  prevalence  mapping exercise around the perimeter of the Majete wildlife reserve, Chikwawa  District, Southern  Malawi through an rMIS that will be conducted monthly over
a two-year period.  Section \ref{sec:discussion} is a concluding  discussion.

\section{Methodological framework}\label{sec:methods}

\subsection{Geostatistical models for prevalence data}

The standard geostatistical model for prevalence data can be formulated as follows \citep*{Diggle1998}. For $i=1,...,n$,
let $Y_i$ be the number of positive outcomes out of $n_i$ individuals tested at location $x_i$ in a region of interest $D \subset \IR^2$, and $d_i \in \IR^p$ a vector of associated covariates. The model assumes
that $Y_i \sim {\rm Binomial}(n_i, p(x_i))$ where $p(x)$ is the prevalence of disease at a location $x$.
The model further assumes that
\begin{equation}
\log[p(x_i)/\{1-p(x_i)\}] = d(x_i)^\prime \beta + S(x_i)
\label{eq:prevalence}
\end{equation}
where $S(x)$ is a stationary Gaussian process with zero mean, variance $\sigma^2$ and correlation function $\rho(u) = {\rm Corr}\{(S(x),S(x^\prime)\}$, where $u$ is the distance between $x$ and $x^\prime$.

Fitting the standard model involves computationally intensive Monte Carlo methods, but software implementations are available; we use the {\tt R} package {\tt PrevMap} \citep{Giorgi2015}.  \citet{Stanton2013} show that provided the $n_i$ are at least 100 and $|p(x)-0.5|$ is at most 0.4, reliable predictions can be obtained using the following computationally simpler approach. Define the {\it empirical logit transform}, $$Y_i^* = \log\{(Y_i+0.5)/(n_i - Y_i + 0.5)\}$$ and assume that

\begin{equation}
Y_i^* = d(x_i)^\prime \beta + S(x_i) + Z_i,
\label{eq:normal_approx}
\end{equation}

where, as in (\ref{eq:prevalence}), $S(x)$ is a stationary Gaussian process with variance $\sigma^2$ and correlation function $\rho(u)$, and the $Z_i$ are mutually independent zero-mean Gaussian random variables with variance $\tau^2$.  Using this approximate method, predictive inferences need to be back-transformed from the logit to the prevalence scale.

In what follows, we will assume a \citet{Matern1960} correlation structure for $S(x)$,

\begin{equation}
\rho(u; \phi; \kappa) = \{2^{\kappa - 1}\Gamma(\kappa)\}^{-1}(u/\phi)^{\kappa}K_{\kappa}(u/\phi),
\label{eq:matern}
\end{equation}

where $\phi > 0$ is a scale parameter that controls the rate at which correlation decays with increasing distance, $K_{\kappa}(\cdot)$ is a modified Bessel function of order $\kappa > 0$, and $S(x)$ is $m$ times mean-square differentiable if $\kappa>m$. In the simulation studies reported in Section \ref{sec:simulation} we use the computationally simpler, approximate method to compare different designs and do not include covariates. For the analyses of the Majete data reported
in Section \ref{sec:application} we use the standard model (\ref{eq:prevalence}).

\subsection{Likelihood-based inference under adaptive design}\label{subsec:likelihood}

Almost all geostatistical analyses are conducted under the assumption that the sampling design, $X$, is stochastically independent of $S$. This justifies basing inference on the likelihood function corresponding  to the conditional distribution of $Y$ given $X$, which typically gives information on all quantities of interest. \citet*{Diggle2010d} discuss the inferential challenges that result when the independence assumption does not hold,  in which case the data $(X,Y)$ should strictly be considered jointly as a realisation of  a marked point process. \citet*{Diggle2010d} call this {\it preferential sampling};  see also \citet*{Pati2011}, \citet*{Gelfand2012b}, \citet*{Shaddick2014}, and Zidek, Shaddick and Taylor (2014) \nocite{Zidek2014}.

In adaptive design, $X$ and $S$ are not independent but are conditionally independent given $Y$, which simplifies the form of the likelihood function. To see why, let $X_0$ denote an initial sampling design chosen independently of $S$, and $Y_0$ the resulting measurement data. Similarly denote by $X_1$ the set of  additional sampling locations added as a result of analysing the initial data-set $(X_0,Y_0)$, $Y_1$ the resulting additional measurement data, and so on. After $k$ additions, the complete data-set consists of $X= X_0 \cup X_1 \cup ... \cup X_k$ and $Y = (Y_0,Y_1,...,Y_k)$. Using the notation $[\cdot]$ to mean ``the distribution of'', the associated likelihood for the complete data-set is

\begin{equation}
\label{eq:integral}
[X,Y] = \int_S [X,Y,S] dS.
\end{equation}

We consider first the case $k=1$. The standard factorisation of any multivariate distribution gives

\begin{equation}
[X,Y,S] = [S,X_0,Y_0,X_1,Y_1] = [S] [X_0|S] [Y_0|X_0,S] [X_1|Y_0,X_0,S] [Y_1|X_1,Y_0,X_0,S].
\label{eq:decomposition}
\end{equation}

On the right-hand side of (\ref{eq:decomposition}), note that by construction, $[X_0|S] = [X_0]$ and $[X_1|Y_0,X_0,S] = [X_1|Y_0,X_0]$. It then follows from (\ref{eq:integral}) and (\ref{eq:decomposition}) that

\begin{eqnarray}
[X,Y] & = & [X_0][X_1|X_0,Y_0] \times \int_S [Y_0|X_0,S] [Y_1|X_1,Y_0,X_0,S] [S] dS \nonumber \\
      & = & [X|Y_0] \times [Y|X].
\label{eq:product}
\end{eqnarray}

This shows that the conditional likelihood, $[Y|X]$, can legitimately be used for inference although, depending on how $[X|Y_0]$ is specified, it may be inefficient.  The argument leading to (\ref{eq:product}) extends to $k>1$ with essentially only notational changes.

\section{An adaptive design strategy}\label{sec:framework}

\subsection{Performance criteria}\label{subsec:performance}

In practice, each geostatistical prediction exercise will have its own, context-specific primary objective. To provide a framework for a general discussion, let $S=\{S(x): x \in D\}$ denote the realisation of the process $S(x)$ over  $D$. Also, let $Y$ denote the data obtained from the sampling design $X=\{x_1,...,x_n\}$,  and $Y=(Y_1,...,Y_n)$ the corresponding measurement data. Denote by $T = {\cal T}(S)$, called the {\it predictive target}, represent the property of $S$ that is of primary interest. A generic measure of the predictive accuracy of a design $X$ is its mean square error, $MSE(X) = {\rm E}[(T - \hat{T})^2]$, where $\hat{T} = {\rm E}[T|Y;X]$ is the minimum mean square error predictor of $T$ for any given design $X$. Note that in the expression for $MSE(X)$  the expectation is with respect to both $S$ and $Y$, whereas in the expression for $\hat{T}$ it is with respect to $S$ holding $Y$ fixed at its observed value.

One obvious predictive target is $T(x) = S(x)$ for an arbitrary location $x \in D$. Another, which may be more relevant when the practical goal is to decide whether or not to launch a public health intervention, is a complete map $T(x) = I(S(x)>c)$, where $I(\cdot)$ is the indicator function and $c$ is a policy-relevant threshold; see, for example,  Figure 3 of \citet*{Zoure2014}. Spatially neutral versions of these targets can be defined by integration over $D$, hence
$$IMSE(X) = \int_D {\rm E}[(T(x) - \hat{T}(x))^2] dx.$$

We emphasise that in any particular application, other measures of performance may be more appropriate. However, for a comparative evaluation of different general design strategies, we adopt $IMSE(X)$ as a sensible generic measure.

\subsection{Some non-adaptive geostatistical designs}\label{subsec:non-adaptive}

Two standard non-adaptive  designs are a {\it completely random} design, in which the sample locations $x_i$ form an independent random sample from the uniform distribution on $D$, and a {\it completely regular} design in which the $x_i$ form a regular square or, less commonly,  triangular  lattice. Geostatistical design problems can be classified according to whether the primary objective is parameter estimation or spatial prediction and, in the latter case, whether model parameters are assumed known or unknown. Our focus is on design for efficient prediction when model parameters are unknown, this being the ultimate goal of most geostatistical analyses. Completely regular designs typically give efficient prediction when the target is the spatial average of $S(x)$, i.e. $T = \int_D S(x) dx$, and model parameters are known; see, for example, \citet[Chapter~5]{Matern1960}. When parameters are unknown, less regular designs have been shown  to be preferable in particular settings see, for example, \citet{Diggle2006}, although a general theory of optimal geostatistical design is lacking.

Most of the previous research on design considerations for prediction assumes a known covariance structure for the data, see, for example, \citet{Benhenni1992, Muller2005} and \citet{Ritter1996}. \citet{Su1993} address the problem of estimating parameters from a random process with a finite number of observations, and measure the design performance by integrated mean square error. They show that random designs are asymptotically optimal. \citet*{Mcbratney1981} address the problem of choosing the spacing of a regular rectangular or triangular lattice design to achieve an acceptable value of the maximum of the prediction variance over the region of interest.  \citet*{Yfantis1987} compare three regular sampling designs, namely the square,  equilateral triangle and regular hexagonal lattices. They conclude that the hexagonal design is the best when the nugget effect is large and the sampling density is sparse.

\citet{Royle1998} and \citet{Nychka1998}  use a geometrical approach that does not depend on the covariance structure  of the underlying process $S(x)$. In this approach, sample points are located in a way that minimises a criterion that is a function of the distances between sampled and non-sampled locations.  \citet{Royle1998} show that the resulting \textit{space-filling} designs generally perform well.

In contrast to the spatial designs for efficient prediction reviewed above, \citet{Zhu2005} consider  designing for efficient covariance structure estimation. They assume the Gaussian model (\ref{eq:normal_approx}) without covariates. Their design criterion is $$V_0(S; \theta) = - \rm log \,\, \rm det \,\,{\cal I}(\theta, S)$$ where ${\cal I}(\theta)$ is the information matrix of the covariance parameters. This is equivalent to $D-$optimality in the context of a linear model with uncorrelated measurement errors. \citet{Russo1984}, \citet{Muller1999}, and \citet{Bogaert1999} consider variogram-based, rather than likelihood-based, parameter estimation. The variogram of $S(x)$ is the function $\gamma(u) = \frac{1}{2}{\rm Var}\{S(x) - S(x^{\prime})\}$ where $u$ is the distance between $x$ and $x^\prime$. \citet{Muller1999} regard a design as optimal if it minimises a suitable measure of the ``size'' of the covariance matrix of the resulting parameter estimates.

More often than not, it is desirable to have designs that compromise between the two analysis objectives of parameter estimation and spatial prediction. Usually, the same dataset is used for covariance structure estimation and prediction of $S(x)$ at unsampled locations. \citet{Zhu2006} address the problem of spatial sampling design for prediction of stationary isotropic Gaussian processes with estimated parameters of the covariance structure. They employ a two-step algorithm that uses an initial set of locations $X_0$ to find the best design for prediction with known covariance parameters and then, conditional on $X_0$, uses the rest to find the best design for estimation of those covariance parameters. \citet{Pilz2006} address a similar design problem but using a model-based approach in choosing an optimal design for spatial prediction in the presence of uncertainty in the covariance structure. Using a Bayesian approach, \citet{Diggle2006} consider designs that are efficient for spatial prediction when parameters are unknown. They looked at two different design scenarios, namely: \textit{retrospective} design, where they use as performance criterion the average prediction variance (APV), 	
\begin{equation}
APV = \int_D \rm {Var}\{\textit{S(x)}|\textit{Y}\}d\textit{x},
\label{eq:vbar}
\end{equation}

and \textit{prospective} design, with performance criterion the expectation of APV, with respect to the process $S(x)$. They concluded that in either situation, inclusion of close pairs in an otherwise regular  lattice design is generally a good choice.

\subsection{A class of adaptive designs} \label{sec:design}

Our proposed approach to adaptive geostatistical design is as follows.
\begin{enumerate}
\item Specify the finite set, $X^*$ say,  of $n^*$ potential sampling locations $x_i \in D$. In our motivating application, this consists of the locations of all households in their respective villages in the Majete perimeter area. In other applications, any point $x \in D$ may be a potential sampling location, in which case we take $X^*$ to be a finely spaced regular lattice to cover $D$.
\item Use a non-adaptive design to choose an initial set of sample locations, $X_0=\{x_i \in D: i=1,...,n_0\}$.
\item Use the corresponding data $Y_0$ to estimate the parameters of an assumed geostatistical model.
\item  Specify a criterion  for the addition of one or more new sample locations to form an enlarged set $X_0 \cup X_1$. A simple example  would be for $X_1$ to be the elements of $X^*$ with the largest values of the prediction variance amongst all points not already included in $X_0$.
\item Repeat steps 3 and 4 with augmented data $Y_1$ at the points in $X_1$.
\item Stop  when  the required number of points has been sampled, a required performance criterion has been achieved or no more potential sampling points are
available.
\end{enumerate}

Within this general approach, in addition to choosing a suitable addition criterion in step 4, we need to choose the number and locations of points in the initial design, $X_0$, and the number to be added at each subsequent stage, called the {\it batch size}. A batch size $b = 1$ must be optimal theoretically, but is often infeasible in practice. For example, in our application to prevalence mapping in the Majete wildlife reserve perimeter area, the associated sampling involves field work in challenging terrain and remote villages to obtain the measurements $Y$. Restricting each field-trip to collection of a single measurement would be a hopelessly inefficient use of limited resources.

\subsection{Types of adaptive designs}\label{subsec:types}
We develop two main types of adaptive geostatistical designs namely: \textit{singleton} and \textit{batch} adaptive designs.

In \textit{singleton adaptive sampling}, $b = 1$, i.e. locations are chosen sequentially, allowing $x_{k+1}$ to depend on data obtained at all earlier locations $x_1,\ldots,x_{k}$. In singleton adaptive sampling, one possible addition criterion is to choose $x_{k+1}$ to be the location $x$ with the largest prediction variance of $S(x)$ given the data from  $x_1,\ldots,x_k$. This is an example of a deterministic rule for identifying and adding new sample locations.

In {\it batch adaptive sampling}, $b>1$. A naive extension of the above addition criterion, choosing $(x_{k+1},...,x_{k+b})$ to be the $b$ available locations with the largest prediction variance of $S(x)$, is likely to fail because it does not penalise sampling from multiple locations $x$ at which the corresponding $S(x)$ are highly correlated.

\subsection{Algorithm for adaptive geostatistical design}\label{subsec:algorithm}

For a predictive target $T(x) = S(x)$, given an initial set of sampling locations $X_0= (x_1,...,x_{n_0})$, the available set of additional sampling locations is $A_0 = X^* \setminus X_0$. For each $x \in A_0$, denote by $PV(x)$ the prediction variance,
${\rm Var}(T|Y_0)$. For the Gaussian model (\ref{eq:normal_approx}), $$PV(x) = \sigma^2(1 - r^{\prime}V^{-1}r),$$ where $r = (r_1,\ldots,r_{n_0})$ with $V = \sigma^2R + \tau^2I$, $R$ is the $n\,\, {\rm by} \,\, n$ matrix with elements $r_{ij} = \rho(||x_i - x_j||)$ and $I$ is the identity matrix.

We propose to incorporate a \textit{minimum distance} addition criterion, whereby we choose new locations $x_{{n_0}+1},x_{{n_0}+2},...,x_{{n_0}+b}$ with the $b$  largest values of $PV(x)$ subject to the constraint that no two locations are separated by a distance of less than $\delta$.

For a formal specification, we use the following notation:\\
    \hspace*{0.2in}$\cdot \;$ $X^*$ is the set of all potential sampling locations, with number of elements of $n^*$;\\
    \hspace*{0.2in}$\cdot \;$ $X_0$ is the initial sample, with number of elements $n_0$;\\
    \hspace*{0.2in}$\cdot \;$ $b$ is the batch size;\\
    \hspace*{0.2in}$\cdot \;$ $n = n_0 + kb$ is the total sample size;\\
    \hspace*{0.2in}$\cdot \;$ $X_j, j \geq 1,$ is the set of locations added in the $j^{th}$ batch, with number of elements $b$;\\
    \hspace*{0.2in}$\cdot \;$ $A_j = X^* \setminus (X_0 \cup ...\cup X_j)$ is the set of available locations after addition of the $j^{th}$ batch.

The algorithm then proceeds as follows.
\vspace{-0.05in}
\begin{enumerate}
\item Use a non-adaptive design to determine $X_0$.

\item Set j=0

\item For each $x \in A_j$, calculate $PV(x)$: \\
   \hspace*{0.2in} (i) choose $x^* = {\rm arg \,\, max}_{A_j} PV(x)$,\\
   \hspace*{0.2in} (ii) if $||x^*-x_i|| > \delta$, for all $i=1,...,n_0 + jb$, add $x^*$ to the design,\\
   \hspace*{0.2in} (iii) otherwise, remove $x^*$ from $A_j$

\item Repeat step 3 until $b$ locations have been added to form the set $X_{j+1}$.

\item Set $A_j = A_{j=1} \setminus X_j$ and we update $j$ to $j+1$.

\item Repeat steps 3 to 5 until the total number of sampled locations is $n$ or
$A_j = \emptyset$.

\end{enumerate}

\section{Simulation study}\label{sec:simulation}

We conducted a simulation study of adaptive geostatistical design (henceforth AGD) so as to compare its performance with standard examples of non-adaptive geostatistical designs (NAGD). Sampling in non-adaptive designs is based on {\it a priori} information and is fixed before the study is implemented \cite{Thompson2002}. Two examples of NAGD are: \textit{random} and \textit{inhibitory} design. Inhibitory designs use a constrained form of simple random sampling \cite{Diggle2013} whereby the distance between any two sampled locations is required to be at least $\delta$. In this way, we  retain the objective of a randomised design whilst guaranteeing a relatively even spatial coverage of the study region.

In each case, data were generated as a realisation of Gaussian process $S(x)$ on a 64 by 64 grid covering the unit square, giving a total  of $n^*=4096$ potential sampling locations. We specified $S(x)$ to have expectation $\mu = 0$, variance $\sigma^2 =1$ and Mat\'{e}rn correlation function (\ref{eq:matern}), with  $\phi$ = 0.05 and $\kappa$ = 1.5, and no measurement error, i.e. $\tau^2$ = 0. In each run of the simulation, we used the adaptive design algorithm outlined in Section \ref{subsec:algorithm} to sample a total of $n = 100$ locations.  We varied the initial sample size $n_0$ between 30 and 90  and considered batch sizes $b = 1$ (singleton adaptive sampling), 5 and 10.

\subsection{Adaptive vs non-adaptive sampling}
\begin{figure}[H]
    \centering
    \begin{subfigure}[b]{0.48\textwidth}
        \centering
        \includegraphics[width=\textwidth]{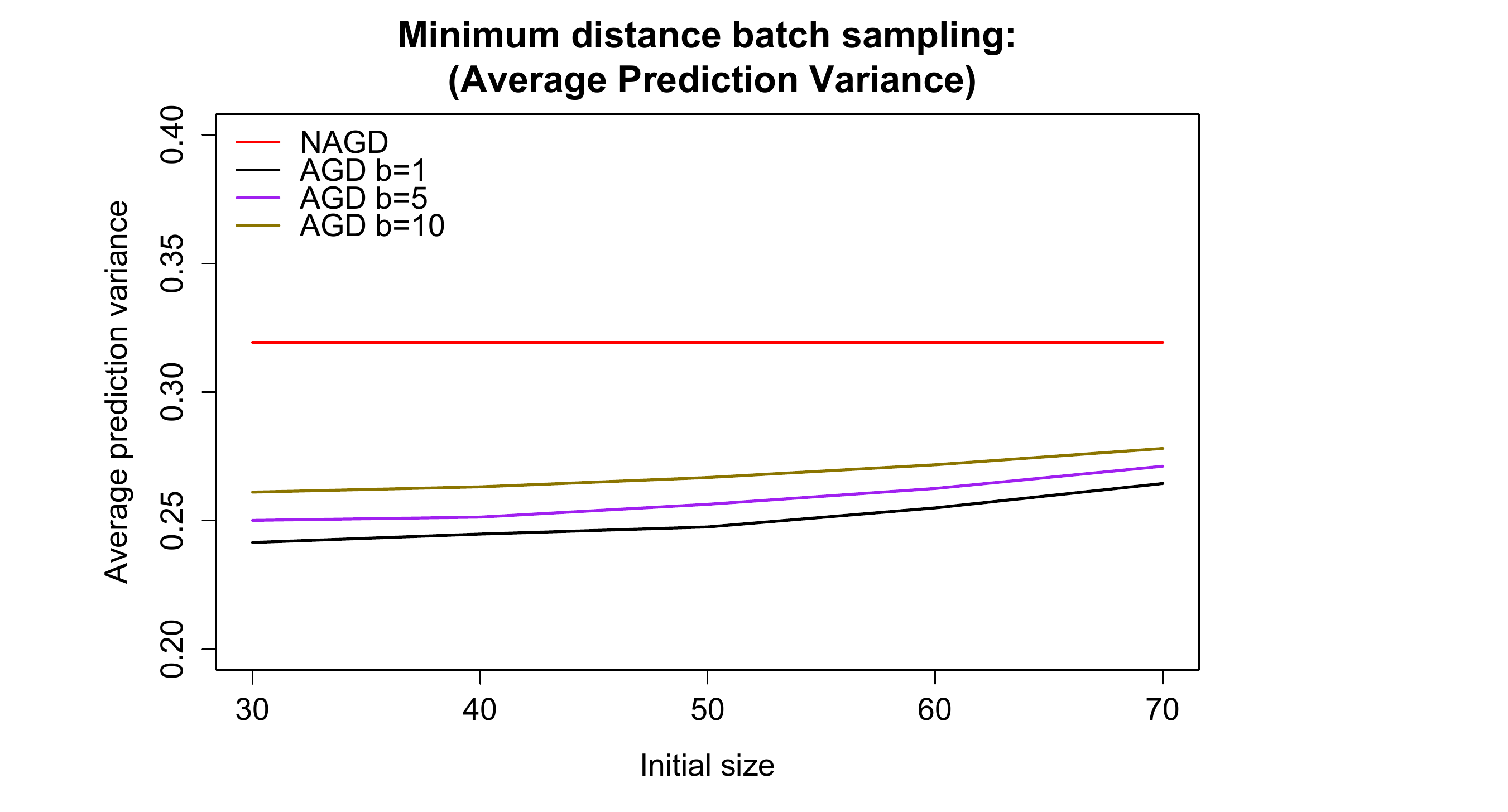}
        \caption{$n_0 = 30$}
        \label{fig:comparison1}
    \end{subfigure}
    \hfill
    \begin{subfigure}[b]{0.48\textwidth}
        \centering
        \includegraphics[width=\textwidth]{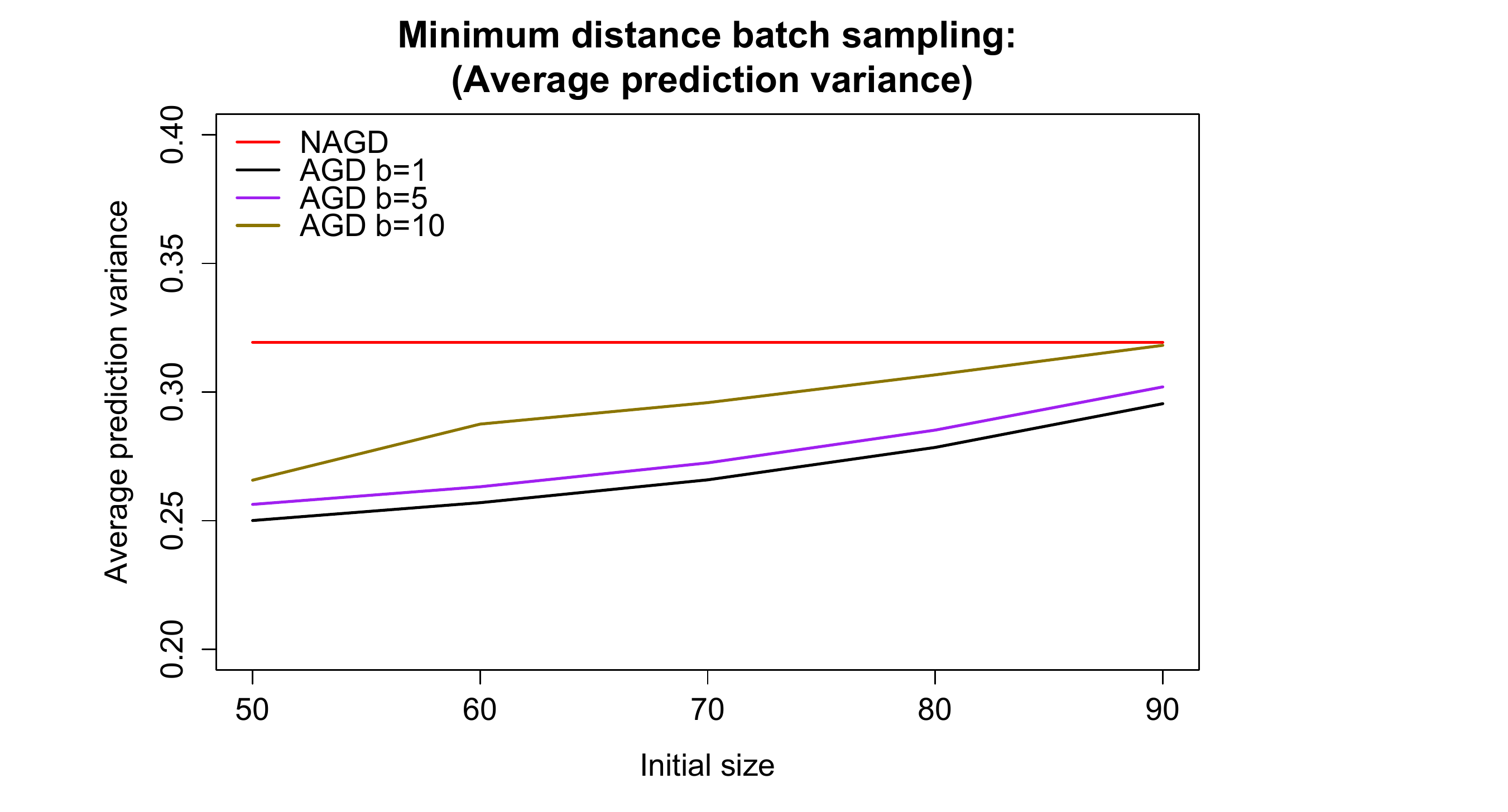}
        \caption{$n_0 = 50$}
        \label{fig:comparison2}
    \end{subfigure}
		\caption{Non-adaptive (NAGD) vs minimum distance batch adaptive (AGD) sampling,
   with $\delta=0.03$ and AGD batch sizes $b=1, 5$ and 10. In the left-hand panel, $n_0=30$;
    in the right-hand panel, $n_0=50$. See text for details of the simulation model.}
	\label{fig:comparison}
\end{figure}

For the non-adaptive sampling of each realisation, and for the initial sample in adaptive
sampling,
we used an inhibitory design with $\delta = 0.03$.
We evaluated each design by its spatially averaged prediction variance, i.e. APV as defined at (\ref{eq:vbar}), in turn averaged over 100 replicate simulations. When the initial sample size is $n_0=30$, the left-hand panel of Figure \ref{fig:comparison} shows singleton adaptive sampling to have the lowest APV, achieving a value APV = 0.24. As the size of the batch increases, APV also increases, but remains substantially lower than the value  APV=0.33 achieved by non-adaptive sampling.

As the initial size $n_0$ increases towards $n=100$, the APV for any of the AGDs necessarily approaches that of the NAGD. For example, the right-hand panel of Figure \ref{fig:comparison} shows the results when $n_0=50$. The value of APV $\approx$ 0.33 when $n_0=90$ and $b = 10$. For $b = 1$ and 5, APV generally remains low whilst steadily approaching that of NAGD when $n_0$ increases towards $n$.

\section{Application: rolling malaria indicator surveys for \\ malaria prevalence in the Majete perimeter}\label{sec:application}

In this Section, we illustrate the use of our proposed sampling methodology to construct a malaria prevalence map for part of an area of the community surrounding Majete wildlife reserve within Chikwawa district (16$^{\circ}$ 1$^{\prime}$ S; 34 $^{\circ}$ 47$^{\prime}$ E), in the lower Shire valley, southern Malawi. The Shire river (the biggest river in Malawi) runs throughout the length of Chikwawa district, causing perennial flooding in the rainy season. Chikwawa is situated in a tropical climate zone with a mean annual temperature of 26 $^{\circ}$C, a single rainy season from November to April and annual rainfall of approximately 770 mm. The district has extensive rice and sugar-cane irrigation schemes.

The area surrounding Majete wildlife reserve forms the region for a five-year monitoring and evaluation study of malaria prevalence, with an embedded randomised trial of community-level interventions intended to reduce malaria transmission. The whole Majete perimeter is home to a population of $\approx$ 100,000. Within this population, three distinct administrative units known as focal areas A, B and C have been selected to form the study region. These are spread over 61 villages with $\approx$ 6,600 households and a population of $\approx$ 24,500. Here, we illustrate adaptive sampling design methodology using data from focal area B, see Figure \ref{fig:majete_map}.

The first stage in the geostatistical design was a complete enumeration of households in the entire study region, including their geo-location collected using Global Positioning System (GPS) devices on a Samsung Galaxy Tab 3 running Android 4.1 Jellybean operating system. These devices are accurate to within 5 meters.  In the on-going rMIS, approximately 90 households are sampled per month per focal area, so that each household will be visited twice over the two years of the study. Malaria prevalence is highly seasonal. The adaptive design problem therefore consists of deciding which households to sample in each of the first 12 months so as to optimise the precision of the resulting sequence of 12 prevalence maps. In year 2, the sampling design will be re-visited to take account of both statistical considerations and any practical obstacles encountered during the first year. Here, to illustrate the methodology, we use data from the first wave of sampling.

\subsection{Data}\label{data}
\begin{figure}[!t]
	\centering
		\includegraphics[width = 10cm, height = 10cm]{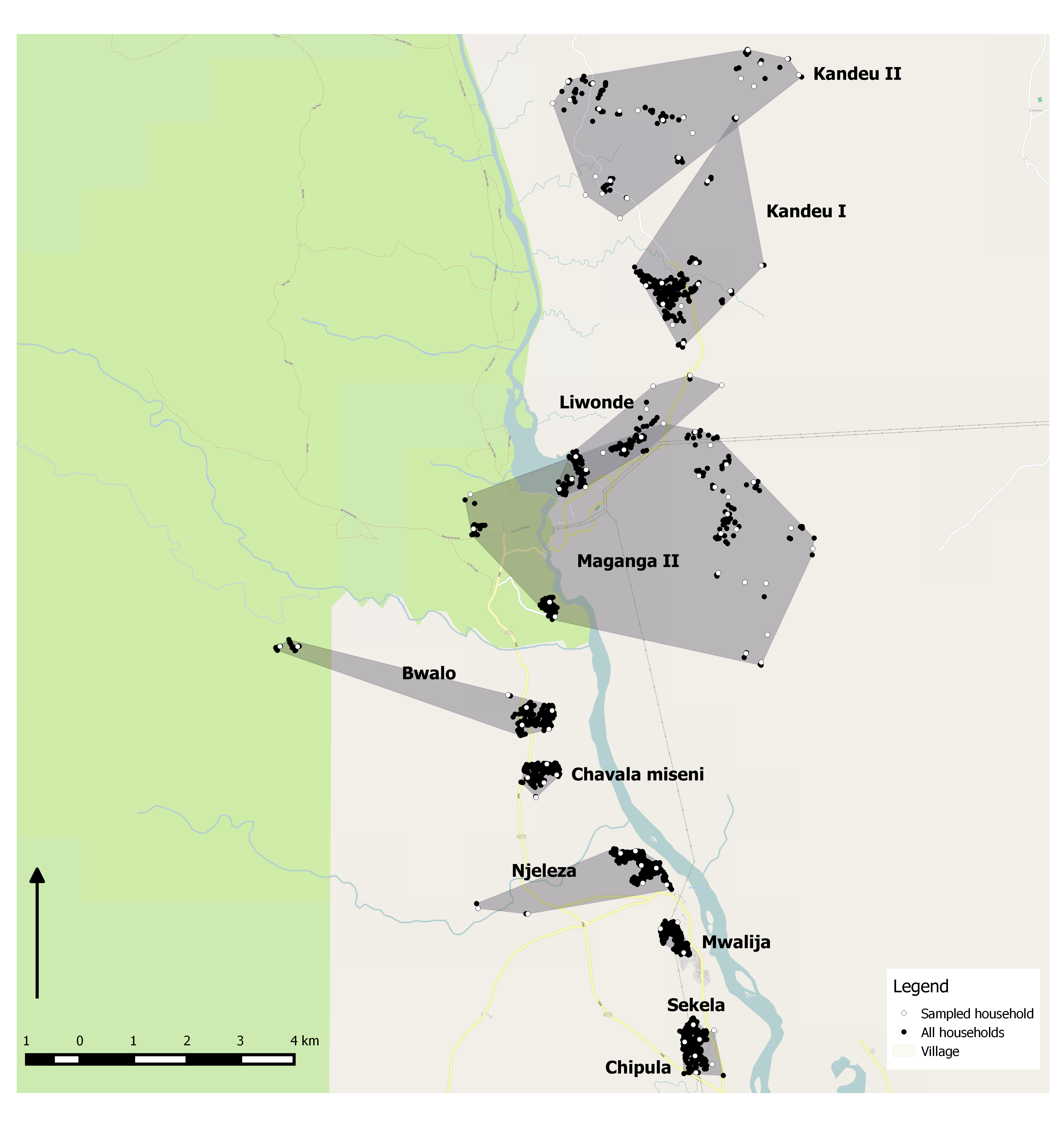}
	\caption{Households within the Majete wildlife reserve perimeter in focal area B (black dots) and sampled household locations (white dots) shown in their respective villages.}
	\label{fig:majete_map}
\end{figure}

The initial population-level continuous malaria indicator survey was conducted over the period April to June 2015. The survey recruited children aged less than 5 years and women of child bearing age, 15 to 49 years, in 10 village communities in order to monitor the burden of malaria.  An inhibitory sampling design was used to sample an initial 100 households per focal area. Selection of the households was as  follows. Households were randomly selected within each village from a list of enumerated households, whilst ensuring a good spatial coverage of the focal area by insisting that the distance between any two sampled households is not less than 0.1 kilometres. Figure \ref{fig:majete_map} shows the sampled household locations (red dots) in their respective villages, with black dots indicating all households in each village. Data collected from the target population include individual level outcomes of a malaria rapid diagnostic test and covariates including age and gender. Household level covariates such as socioeconomic status and household location were also collected.

For predictive mapping, any covariates included in the model must be available at all prediction locations. We used two digital elevation model (DEM) derivatives, elevation and normalized difference vegetation index (NDVI), which are readily available throughout the study region. Data for these covariates were derived using the Advanced Space-borne Thermal Emission and Reflection Radiometer (ASTER) Global DEM version 2. ASTER GDEM V2 has a spatial resolution of 30 meters. The data were downloaded from the United States Geological Survey (USGS) through their `Global Data Explorer' \url{http://gdex.cr.usgs.gov/gdex/}.

\subsection{Results}\label{results}
We emphasise that at this  early stage of the Majete study the data are too sparse for a definitive prevalence analysis but sufficient for adaptive sampling design methodology illustration. The response from each individual in a sampled household is the binary outcome of a rapid diagnostic test (RDT) for the presence/absence of malaria from a finger-prick blood sample. Out of the 100 households
in the initial sample, 72 had at least one individual who met the inclusion criteria (see Section \ref{data} above). The total
number of eligible individuals in these 72 households was 126, with household size ranging from 1 to 8 individuals. For covariate selection we used ordinary logistic regression, retaining covariates with nominal $p$-values less than than 0.05. This resulted in the set of covariates shown in Table \ref{EstimTable}, with terms for elevation, NDVI and the interaction between the two.  We then fitted the binomial logistic model (\ref{eq:prevalence}) to obtain the Monte Carlo maximum likelihood estimates of the parameters and associated 95 \% confidence intervals also shown in Table \ref{EstimTable}. Each evaluation of the log-likelihood used 10,000 simulated values, obtained by conditional simulation of 110,000 values and sampling every $10^{th}$ realization after discarding a burn-in of 10,000 values.

\begin{table}[H]
\centering
\begin{tabular}{llll}
\hline
Term & Estimate & \multicolumn{2}{c}{95 \% Confidence Interval} \\
\hline
${\rm Intercept}$ & -5.4827 & \multicolumn{2}{c}{(-7.6760, -3.2893)} \\
${\rm Elevation}$ & 0.02651 & \multicolumn{2}{c}{(0.0162, 0.0368 )} \\
${\rm NDVI}$ & 4.6130 & \multicolumn{2}{c}{(0.1581, 9.0680)} \\
${\rm Elev. \times NDVI}$ & -0.0405 & \multicolumn{2}{c}{(-0.0588, -0.0223)} \\
$\sigma^2$ & 0.6339 & \multicolumn{2}{c}{(0.4438, 0.9055)} \\
$\phi$ & 0.2293 & \multicolumn{2}{c}{(0.1042, 0.5049)} \\
\hline
\end{tabular}
\caption{Monte Carlo maximum likelihood estimates and 95 \% confidence intervals for the model fitted to the Majete malaria data.}
\label{EstimTable}
\end{table}

 From Table \ref{EstimTable}, elevation and NDVI show positive marginal associations with malaria, with a negative interaction. Focal area B is divided through its length by the Shire river.  The north-east part has relatively high elevation and NDVI values. Prevalence is generally low in the south-west of the region, whereas the north-east has pockets of comparatively high malaria prevalence. This suggests that heterogeneity in malaria prevalence over focal area B involves other risk factors (social or environmental) that are not available in the current data.

\begin{figure}[H]
	\centering
		\includegraphics[height = 10cm, width = 10cm]{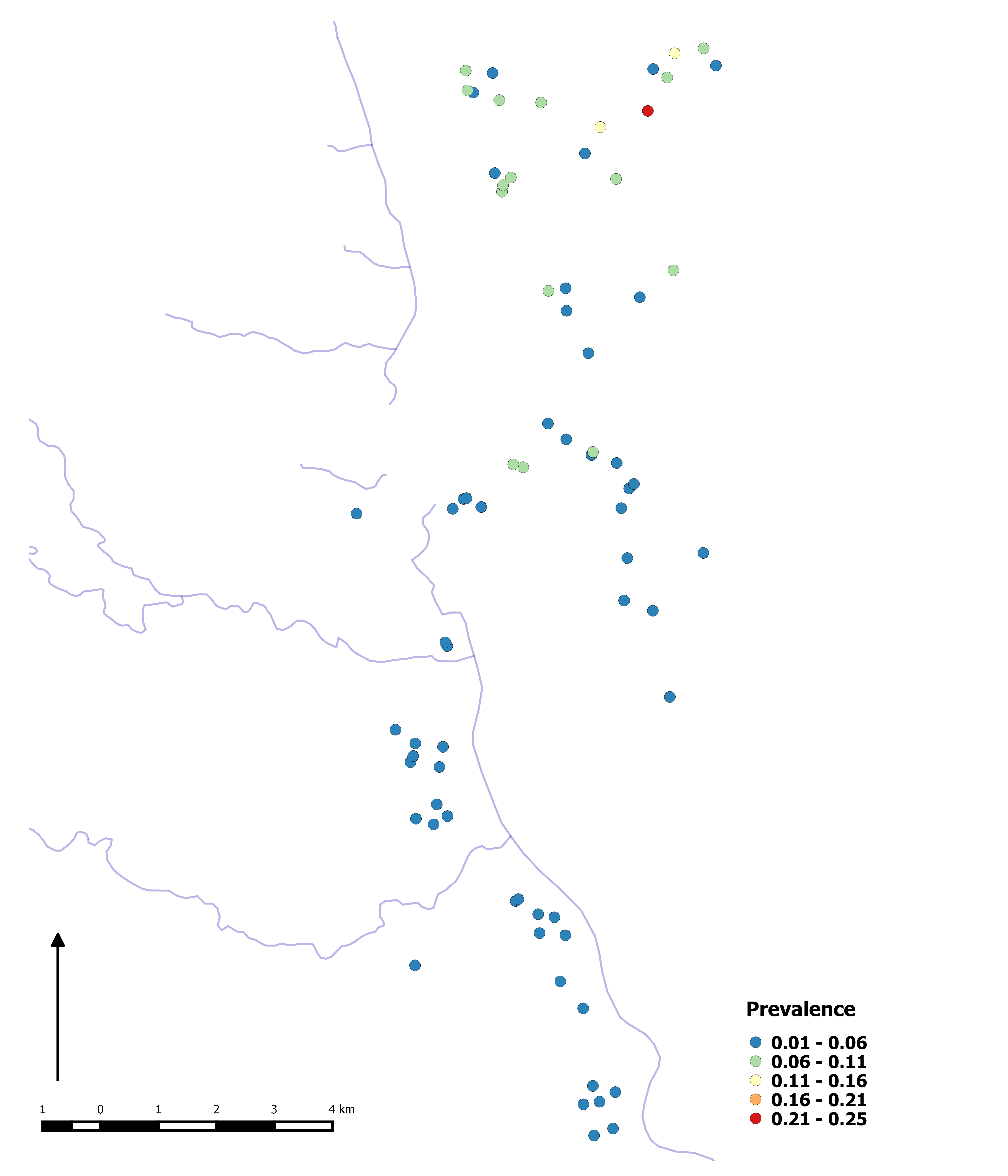}
	\caption{Predictions of $d(x)^{\prime} \beta + S(x) $ at observed locations in focal area B. The blue lines shows Shire and Matope rivers.}
	\label{fig:pointpred}
\end{figure}

Figure \ref{fig:pointpred} shows the predicted prevalence at each of the observed locations. Households at high altitude and under dense vegetation cover have generally high malaria prevalence. For this study, the elevation of households varied from 60 to 460 meters above sea level. Rivers and streams that are fast flowing in nature are not generally favourable for mosquito larvae; the Shire river is a big and fast flowing river. Sampling was done at the time of peak malaria transmission at the end of the rainy season when rains subside. This could potentially explain the low prevalence in the southern part of the study region. Also, the high prevalence area in the north-east is generally more remote with far access to health facilities.

\subsection{Adaptive sampling}
\begin{figure}[H]
    \centering
    \begin{subfigure}[b]{0.483\textwidth}
        \centering
        \includegraphics[width=\textwidth]{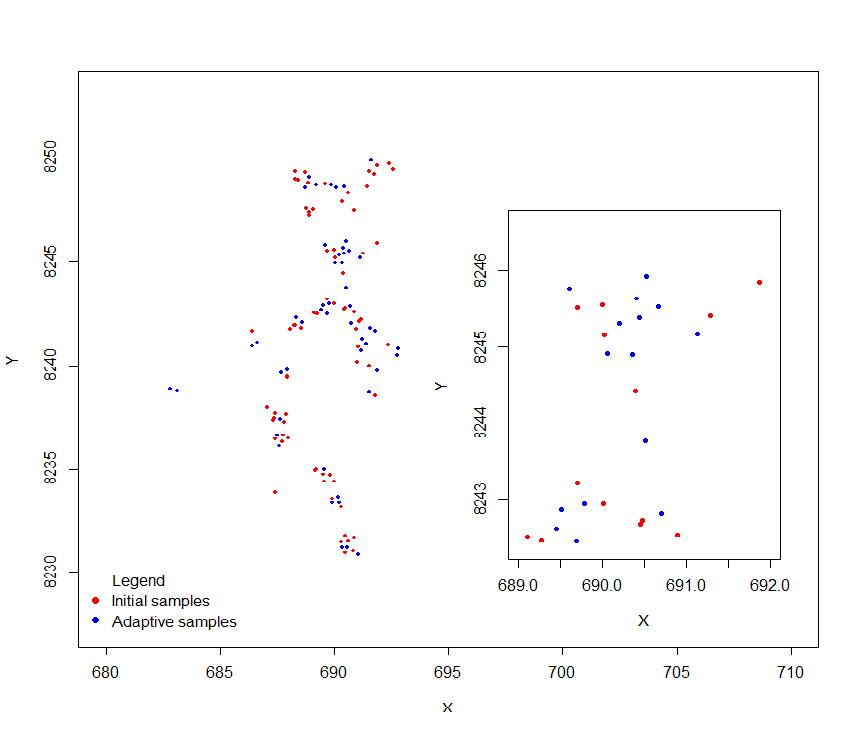}
        \caption{}
        \label{fig:adaptivesample}
    \end{subfigure}
    \hfill
    \begin{subfigure}[b]{0.41\textwidth}
        \centering
        \includegraphics[width=\textwidth]{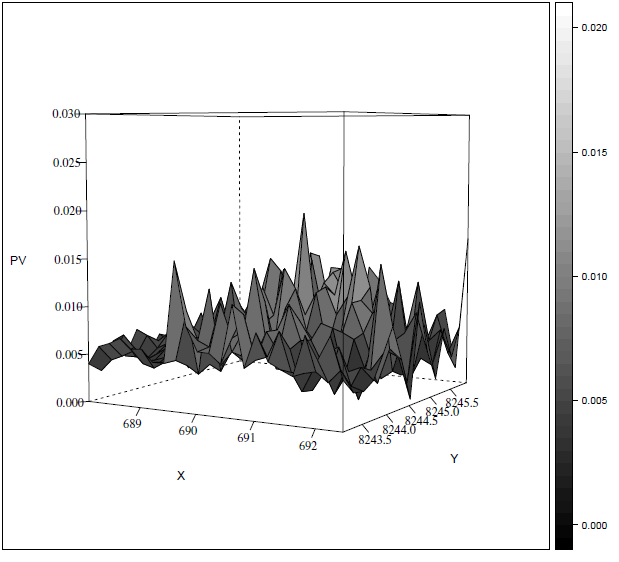}
        \caption{}
        \label{fig:PredictionVarianceSurface}
    \end{subfigure}
		\caption{(a) Initial inhibitory sampling design locations (red dots) and adaptive sampling design locations (blue dots) in focal area B. Inset shows a subset of locations. \hspace{0.1cm} (b) Prediction variance surface for the inset sub-region from \ref{fig:adaptivesample}.}
	\label{fig:pv-samples}
\end{figure}

We now use the \textit{minimum distance batch adaptive sampling} approach explained in Section \ref{subsec:algorithm} to determine new locations that can and should be added to the existing sample in an adaptive manner. We first calculate the prediction variance at each household using the data from the 72 initial sample locations, shown as red dots in Figure \ref{fig:adaptivesample}. Prediction variances range between 0.0003 and 0.0325, and are relatively small at locations closer to the observed locations, although this depends on the number of eligible individuals at each location. We then choose a sample of 50 additional locations using the algorithm outlined in Section \ref{subsec:algorithm} above. The blue dots in Figure \ref{fig:adaptivesample} show these 50 new locations determined using the minimum distance threshold $\delta = 0.15$ kilometres. The new sampling locations are well spread across the study region, which is beneficial for area-wide spatial prediction. Also, although we have imposed $\delta$ between any two sampled locations in order to penalise highly correlated multiple sample locations, the new sample locations nevertheless include some  pairs of old and new locations in which the new location has been chosen to be relatively close to an initial location with high prediction variance; recall that the number of eligible individuals per household varied between 1 and 8, hence the prediction variance at a sampled location is itself highly variable. As noted earlier,  closely spaced pairs are helpful for effective spatial prediction when the true model parameters are not known, which is the reality in most geostatistical problems.

In Figure \ref{fig:PredictionVarianceSurface} we show the prediction variance surface after addition of the 50 adaptively sampled locations, for the sub-region highlighted in Figure \ref{fig:adaptivesample}. Locations with high prediction variance are potential candidates for the next round of adaptive sampling, subject to their meeting the minimum distance constraint.

The adaptive sampling design criterion ensures
 that data are collected only from locations that will deliver useful
 additional information in order to understand the spatial heterogeneity throughout the study region.

\section{Discussion}\label{sec:discussion}
In any particular application, the objectives of the study can and should inform the design strategy. We have developed an adaptive sampling strategy within a model-based geostatistics framework for survey based disease mapping  in poor resource settings. The minimum  distance  batch sampling design described in Section \ref{subsec:algorithm} is intended to deliver efficient mapping of the complete surface, $S(x)$, over the region of interest.  Detection  and  subsequent  evaluation of sub-regions where policy-determined prevalence thresholds could help guide more targeted intervention measures, would require progressive concentration of sampling into areas of relatively high prevalence.

In our application to malaria prevalence mapping, we used an initial set of rMIS data to map disease prevalence in focal area B and  analysed the resulting data to define a follow-up sample of new locations with the aim of reducing as much as possible the average prediction variance.  The batch size is large because of the high cost in staff and travel time of re-visiting the study region more often than monthly. Smaller batch sizes, if feasible, would potentially lead to greater gains in efficiency.

The adaptive sampling design approach is of potentially wide application to disease mapping in low resource settings, where accurate registry data typically do not exist. Mapping exercises are an important component of any control or elimination  programme.
Collecting data adaptively allows for local identification and targeting of areas with high transmission, incidence or prevalence, and an understanding of which household-level and community-level factors influence these properties. Knowledge of these properties can inform area-wide health policymaking and identify locations of greatest need where interventions that would be considered too costly or complicated to implement across an entire population can be targeted in order to optimise their public health impact.

The choice of the initial sampling design $X_0$ is an important step for adaptive sampling. The initial sample size, $n_0$, needs to be large enough to allow the fitting of a geostatistical model, whose estimate parameter values then drive the adaptive sampling. In the Majete application, we prescribed $n_0=100$ but, in the event, found eligible study participants in 72 of the sampled households. We recommend re-estimation of the model parameters after each batch of locations has been added.

In the Majete application, the irregular spatial distribution of households across the study-region meant that the set of 122 sampled locations after the first batch of adaptively sampled locations had been added to the initial design achieved a good compromise between even coverage of the study-region and the inclusion of close pairs, which is generally helpful for efficient parameter estimation.  In other contexts, and specifically where there is essentially no restriction of the placement of sampling locations, it would be better to use an initial design that deliberately includes some close pairs, as recommended in \citet{Diggle2006}.

In conclusion, the proposed adaptive sampling design approach provides a systematic approach to the collection of exposure and outcome data over time  so as to optimise progress towards achievement of the analysis objective. Adaptive designs are particularly well suited to spatial mapping studies in low resource settings where uniformly precise mapping may be unrealistically costly and the priority is often to identify critical areas where interventions can have the greatest health impact. Development of adaptive geostatistical design methodology is therefore timely for monitoring and evaluating interventions in tropical diseases with high burden such as malaria, in areas where accurate disease registries do not exist and resources are severely limited. Malaria in particular is a leading cause of death in most of sub-Saharan Africa, especially among children under 5 years of age. Malaria monitoring and control programmes can benefit from the availability of accurate prevalence maps. Geostatistical analysis in conjunction with adaptive sampling is an effective, practical strategy for producing accurate local-scale maps that can pick up short-term changes in disease burden and that are complementary to the national-scale maps that have been produced, for example, by \citet{Hay2004}, \citet{Guerra2007}, \citet{Hay2009} and \cite{Gething2012}.

\section*{Acknowledgement}
We thank the participants and Majete integrated malaria control project staff involved in the ongoing data collection of the presented household prevalence surveys, part of which is here.

\section*{Funding}
Michael Chipeta is supported by ESRC-NWDTC Ph.D. studentship (grant number ES/J500094/1).\\
Dr Dianne Terlouw, Dr Kamija Phiri and Prof. Peter Diggle are supported by the Majete integrated malaria control project grant funded by the Dioraphte foundation.

\bibliography{library}
\bibliographystyle{myplainnat}

\end{document}